\documentstyle[fleqn]{article}
\newcommand{\bg}{\begin{equation}}
\newcommand{\eg}{\end{equation}}

\def\diag{\mbox{diag }}

\def\id{\mbox{\rm 1\hskip-.25em l}}

\def\LD{{\cal L}}

\def\lplus{\supset\mkern -19mu\hbox{\small +}\mkern 10mu}

\newcommand{\mb}[1]{\mbox{\boldmath $#1$}}

\begin{document}
%% \begin{flushright}
%% MZ-TH/97-29
%% \end{flushright}
%% \bigskip\bigskip
\centerline{\normalsize\bf TRIANGULAR MASS MATRICES OF QUARKS AND}
\baselineskip=16pt
\centerline{\normalsize\bf CABIBBO-KOBAYASHI-MASKAWA MIXING}
\baselineskip=60pt
\centerline{Rainer H\H au\ss ling$^a$ and Florian Scheck$^b$}
\baselineskip=12pt
\begin{center}
{\footnotesize{\it {}$^a$ Institut f\H ur Theoretische Physik,
Universit\H at Leipzig, D-04109 Leipzig,\\
e-mail:\/} {\tt Haeussli@tph204.Physik.Uni-Leipzig.DE}}\\[4pt]
{\footnotesize{\it {}$^b$ Institut f\H ur Physik, Johannes
Gutenberg-Universit\H at, D-55099 Mainz,\\
e-mail:\/} {\tt Scheck@dipmza.Physik.Uni-Mainz.DE}}
\end{center}
\vspace*{3cm}
\abstract{Every nonsingular fermion mass matrix, by an appropriate unitary
transformation of right-chiral fields, is equivalent to a triangular
matrix. Using the freedom in choosing bases of right-chiral fields in the
minimal standard model, reduction to triangular form reduces the well-known
ambiguities in reconstructing a mass matrix to trivial phase redefinitions.
Furthermore, diagonalization of the quark mass sectors can be shifted to one
charge sector only, without loosing the concise and economic triangular
form. The corresponding effective triangular mass matrix is reconstructed,
up to trivial phases, from the moduli of the CKM matrix elements, and vice
versa, in a unique way. A new formula for the parametrization independent
CP-measure in terms of observables is derived and discussed.}

\vfill
\newpage
\normalsize\baselineskip=15pt

\noindent
{\bf 1. Introduction}

Although generation mixing, eventually, must be due to as yet
unknown physics beyond
the minimal standard model, its parametrization in terms
of the Cabibbo-Kobayashi-Maskawa (CKM) matrix is quite restrictive and
could, in fact, turn out to be inconsistent with precision
measurements of weak decays and CP-violating amplitudes. With the
restriction to three generations unitarity of the CKM matrix alone
imposes (nonlinear) constraints on observables which may or may not
be obeyed by experiment, see e.g. \cite{WXI,ChK,Jar,Gre}. Further constraints 
are obtained if the mixing matrix is derived from the primordial mass
matrices of {\it up-\/}type quarks and of {\it down-\/}type quarks.
In adopting this latter strategy and before even invoking specific
models of quark mass matrices, it is important to formulate the mass
terms such that all redundant, unobservable, features are left out from
the start. Only if this is achieved can one hope to sharpen the tests
of compatibility of the CKM scheme for three generations
with experiment. As we shall see below this means much more than the 
well-known eliminating irrelevant phases by
redefinition of basis states.

In this paper we show that the essential information contained in a
given, nonsingular quark mass matrix can be expressed in a particularly
economic and concise way. Making use of the freedom in choosing bases
of right-chiral fields every nonsingular mass matrix is equivalent to a
triangular matrix. We also show that simultaneous diagonalization of
two charge sectors can be shifted to one of them without loosing the
simplicity of triangular matrices. We study examples for the cases
of two and three generations, with and without additional model
assumptions. Implications for invariants that describe CP violating
observables are also discussed.

In sec.~2 we formulate and prove the decomposition theorem that is central
to our analysis. Sec.~3 contains an interpretation of triangular mass
matrices in terms of quark representations and quark mixing. Making use
of the decomposition theorem, we go one step further in sec.~4, and shift
diagonalization to one charge sector only, either to the customary
{\it down\/}-sector, or, equivalently, to the {\it up\/}-sector. This novel
procedure is illustrated by an analytic example with two generations. 
The general case of three generations is treated in sec.~5 which also gives
explicit and analytic formulae for the entries of the CKM matrix.
In sec.~6 we derive and dicuss a new formula for the rephasing invariant 
measure of
CP violation which expresses this quantity in terms of observables only.
In sec.~7 we perform the reconstruction of the effective (triangular) mass
matrix in terms of the elements of the CKM matrix. The final sec.~8 
summarizes our results and offers a few conclusions.

\noindent
{\bf 2. Reduction of nonsingular mass matrices to triangular form}

In order to set the notation we start by recalling a few well-known facts
about the relation of the quark mixing matrix in charged-current (CC)
weak interactions to the mass matrices in the charge $+2/3$ {\it up-\/}quark
sector and the charge $-1/3$ {\it down-\/}quark sector.

The minimal standard model describes CC weak interactions by purely
left-handed currents giving rise to the effective ``V-A'' Lorentz structure
at low energies and maximal parity violation. The right-chiral fields in
the three-generation spinor field
\bg
\Psi =\left(
      u_L^\prime ,d_L^{\,\prime},u_R^\prime,d_R^{\,\prime},
      c_L^\prime ,s_L^\prime ,c_R^\prime ,s_R^\prime,
      t_L^\prime ,b_L^\prime ,t_R^\prime ,b_R^\prime
      \right)^T \label{eq-psi}
\eg
are inert to charged-current interactions and, being singlets with
respect to the weak SU(2) structure group, are fixed only up to
independent, unitary transformations $U_R^{(u)}$ and $U_R^{(d)}$ in the
{\it up-\/} and {\it down-\/}charge sectors, respectively. In
eq.~(\ref{eq-psi}) the primes refer to the weak interaction states, the mass
eigenstates will be denoted by the same symbols without a prime. Thus,
if the mass matrices of {\it up-\/} and {\it down-\/}quarks in the
basis (\ref{eq-psi}) are $M^{(u)}$ and $M^{(d)}$, they are
diagonalized by the bi-unitary transformations
\bg
V_L^{(u)}M^{(u)}V_R^{(u)\,\dagger}\quad\mbox{and}\quad
V_L^{(d)}M^{(d)}V_R^{(d)\,\dagger} \, . \label{eq-biun}
\eg
The Cabibbo-Kobayashi-Maskawa mixing matrix refers to CC interactions,
hence to left-chiral fields only, and is given by the product
\bg
V^{\mbox{\footnotesize (CKM)}}=V_L^{(u)}V_L^{(d)\,\dagger}\, .\label{eq-CKM}
\eg
The unitaries acting on the right-chiral fields, $V_R^{(u)}$ and
$V_R^{(d)}$ remain unobservable and may be chosen at will. We shall make
use of this freedom on several occasions below. In other terms, the
CKM mixing matrix is determined by the mass matrices in the charge
$+2/3$ and $-1/3$ sectors and depends only on the unitary transformations
acting on left-chiral fields, up to the well-known freedom in fixing
the phases of its entries (see e.g. \cite{Sch}).

In this section we show that the essential information that determines
the CKM mixing matrix can be encoded in mass matrices of
{\it triangular form\/}, either upper or lower triangular, provided these
are not singular. For definiteness, in what follows we shall always choose
{\it lower\/} triangle matrices,
\[
T^{(u)}\quad\mbox{and}\quad T^{(d)}\, ,\quad\mbox{with}\quad
T_{ik}^{(u/d)}=0\mbox{ for all }k>i\, .
\]
The importance of this observation for the physics of CC weak interactions
of quarks will be discussed below, in a rather general framework for
generation mixing. The mathematical fact is based on the following lemma
and decomposition theorem \cite{Mur}.\\[2pt]
\noindent
{\bf Lemma}: {\it Let $M$ be an arbitrary, nonsingular matrix of dimension
$n$ and let $H=MM^\dagger$, so that $H$ is a positively definite, hermitean
$n\times n$ matrix. The matrix $H$ can be represented in the form
$H=TT^\dagger$, where $T$ is a nonsingular, lower triangular, matrix of
dimension $n$.}

The lemma is trivially true in dimension 1. For $n>1$ it is proved by
induction with respect to $n$, see \cite{Mur}.

This lemma is used in proving the following theorem:\\[2pt]
\noindent
{\bf Decomposition theorem:} {\it Any nonsingular $n\times n$ matrix
$M$ can be decomposed into the product of a nonsingular, lower triangular
matrix $T$ and a unitary matrix $U$},
\bg
M=TU\, ,\quad \mbox{with}\quad T_{ik}=0\quad\forall\quad k>i\; ,
\quad UU^\dagger =\id \, . \label{eq-MTU}
\eg
{\it This decomposition is unique up to multiplication of $U$ from the left
by a diagonal unitary matrix\/}
$W=\diag (e^{i\omega_1},\ldots ,e^{i\omega_n})$.

Proof: By the lemma the hermitean matrix $H=MM^\dagger$ equals $TT^\dagger$,
with $T$ a lower triangle matrix. This being nonsingular one calculates
$U=T^{-1}M$ and proves $U$ to be unitary: Indeed,
$MM^\dagger =TUU^\dagger T^\dagger =TT^\dagger$. Multiplying by $T^{-1}$
from the left, and by $(T^\dagger )^{-1}$ from the right, $UU^\dagger =\id$.
Suppose now that there is more than one decomposition (\ref{eq-MTU}), say
$M=TU=T^\prime U^\prime$, where $T$ and $T^\prime$ are (lower) triangular.
This means that $T=T^\prime W$, where $W=U^\prime U^\dagger$ is unitary.
The requirement that both $T$ and $T^\prime$ be lower triangular fixes
$W$ to be diagonal. 

Note that the decomposition theorem is equivalent to the Schmidt
orthogonalization procedure. To see this we interpret the rows of
the matrix $M=\{ m_{ik}\}$ as $n$ linearly independent vectors
$\mb{v}^{(i)}=(m_{i1},\ldots ,m_{in}),\;i=1,\ldots ,n$. Likewise,
the rows of $U=\{ u_{ik}\}$ form a set of orthogonal unit vectors
$\mb{u}^{(i)}=(u_{i1},\ldots ,u_{in}),\; i=1,\ldots ,n$.
Equation (\ref{eq-MTU}) then reads
\[
\mb{v}^{(i)}=\sum_{k=1}^{i}T_{ik}\mb{u}^{(k)}\; ,
\quad i=1,\ldots ,n\, .\]

These latter equations are useful in determining the matrix elements
$\{ t_{ik}\}$ of $T$ from those of $M=\{ m_{ik}\}$. For example, in the
case $n=3$ we have
\bg
\vert t_{11}\vert^2 = \sum_{k=1}^3\vert m_{1k}\vert^2 \; .
\label{eq-t1}\eg
\bg
t_{21}=\sum_{k=1}^3m_{1k}^\star m_{2k}/t_{11}^\star\; ,\quad
\vert t_{22}\vert^2 = \sum_{k=1}^3\vert m_{2k}\vert^2-
\vert t_{21}\vert^2 \; .\label{eq-t2}
\eg
\[
t_{31}=\sum_{k=1}^3m_{1k}^\star m_{3k}/t_{11}^\star\; ,\quad
t_{32}=\left[\sum_{k=1}^3m_{2k}^\star m_{3k}-t_{31}t_{21}^\star
\right] /t_{22}^\star\; ,
\]
\bg
\vert t_{33}\vert^2 = \sum_{k=1}^3\vert m_{3k}\vert^2-\vert t_{31}\vert^2-
\vert t_{32}\vert^2\; .\label{eq-t3}
\eg
Note that eqs.~(\ref{eq-t1})--(\ref{eq-t3}) reflect the non-uniqueness
of the unitary $U$ in eq.~(\ref{eq-MTU}) noted above.

In applying the decomposition theorem (\ref{eq-MTU}) to the mass matrix
of the quark sector with charge $q$,
$q=+\frac{2}{3}\mbox{ or }-\frac{1}{3}$, let
$\widehat{M}^{(q)}$ be such that it connects right-chiral fields to conjugate
left-chiral fields. The mass terms in the Lagrangian have the form
\[
\LD_{\mbox{\footnotesize mass}}=
    \sum_{q}\overline{\Psi}\;\widehat{M}^{(q)}\,\Psi +\mbox{ h.c. }\equiv
    \sum_{q}\overline{(\Psi )_L}\;\widehat{M}^{(q)}\, (\Psi )_R +
    \mbox{ h.c. } \, .
\]
Now, replacing $\widehat{M}^{(q)}$ by the product on the r.h.s. of
eq.~(\ref{eq-MTU}), the unitary matrices are absorbed by a
redefinition of the right-chiral fields
\[
\{ u_R^{\prime (n)},\, n=1,2,3\}\equiv
\{ u_R^\prime ,c_R^\prime ,t_R^\prime \}\, ,\;
\{ d_R^{\,\prime (n)},\, n=1,2,3\}\equiv
\{ d_R^{\,\prime} ,s_R^\prime ,b_R^\prime \}\, ,
\]
so that the general mass Lagrangian becomes
\bg
\LD_{\mbox{\footnotesize mass}}=
     \sum_{n,m=1}^3\overline{u_L^{\prime (n)}}\,
     T_{nm}^{(u)}u_R^{\prime (m)}+
     \sum_{n,m=1}^3\overline{d_L^{\,\prime (n)}}\,
     T_{nm}^{(d)}d_R^{\,\prime (m)}+
     \mbox{ h.c. }\, ,\label{eq-mass}
\eg
where $T^{(u)}$ and $T^{(d)}$ are $3\times 3$ (lower) triangular.\\[12pt]
{\bf 3. Interpretation in terms of generation mixing}

Having shown that the essential information on mass matrices in a given
charge sector, by the decomposition theorem, is fully contained in their
triangle form, we now pause to interpret this result in terms of the
physics of generation mixing. For the sake of simplicity let us consider
the case of two generations which, in fact, need not be identical replicas of
each other. Let the fermions of the theory fall into two irreducible
representations of the structure group, say $\rho_I$ and $\rho_{II}$,
of dimensions $n_1$ and $n_2$, respectively, and denote the two sets of
basis states which span these representations 
by $\Psi_I$ and $\Psi_{II}$. Let us assume that we are given an operator
$\widehat{T}$ whose representation in the given basis has triangular form, viz.
\[
\widehat{T}=\left(\begin{array}{cc}
       T_{11} & 0 \\ T_{21} & T_{22}
       \end{array}\right)\, ,
\]
where the diagonal blocks
$T_{11}$ and $T_{22}$ are square matrices of dimension $n_1$ and
$n_2$, respectively, and $T_{21}$ is an $n_1\times n_2$ off-diagonal block.
A unitary transformation $\widehat{R}=\diag (R_I,R_{II})$ of the bases will
take the operator $\widehat{T}$ to
\[
\widehat{T}^\prime= \widehat{R}\widehat{T}\widehat{R}^\dagger =
       \left(\begin{array}{cc}
       R_IT_{11}R_{I}^\dagger & 0 \\[2pt]
       R_{II}T_{21}R_I^\dagger & R_{II}T_{22}R_{II}^\dagger
       \end{array}\right)\, .
\]
Thus, the subspace spanned by $\Psi_{II}$ is an invariant subspace, the
subspace $\Psi_I$ is not. In this situation the combined representation
spanned by $\left( \Psi_I ,\Psi_{II} \right)$ is said to be
reducible but indecomposable, and is written as a semi-direct sum
\bg
\rho_I\lplus\rho_{II}\, . \label{eq-semi}
\eg

This framework is completely general. It seems to us the most economic
and natural parametrization of generation mixing. Indeed, after
extracting all unphysical phases, by redefinition of quark fields,
the triangular form contains the essential, minimal information on
the mass matrix. This is particularly important if one wishes to
obtain analytic expressions of the CKM matrix elements in terms of
quark masses and, possibly, a minimal set of parameters. 
An equivalent way of seeing this is to note that the
CKM-Matrix (\ref{eq-CKM}) is unchanged if the quark mass matrices
$M^{(u)}$ and $M^{(d)}$ are both multiplied from the left by an arbitrary
unitary matrix $X$ such that
\[
M^{\prime\, (u)}=X M^{(u)}\, ,\qquad M^{\prime\, (d)}=X M^{(d)}\, .
\]
In the case of general, non-triangular mass matrices these formulae reflect
the nine parameter freedom in reconstructing the mass matrices from the 
CKM-matrix discussed by Kusenko \cite{Kus}. If, on the other hand, the
mass matrices are required to be triangular before and after the 
transformation, then $X$ must be diagonal, its entries being pure phases.
Thus, the reconstruction of triangular mass matrices is unique up to
a choice of (unobservable) phases.

There may, in fact, be good theoretical reasons for assuming the fermions
of the standard model to fall into representations of the type
(\ref{eq-semi}). For example, representations of this type are
characteristic of graded (or super) Lie algebras, cf. \cite{Scheu}.
Clearly, if a multiplet of scalar Higgs fields appears multiplied with
an operator acting on representations of this type and if the electrically
neutral component of the Higgs develops a nonzero vacuum expectation value, 
fermions of the same charge belonging to different subspaces, via their
Yukawa couplings, acquire triangular mass matrices.
  
In the case of replication of generations the two
terms in (\ref{eq-semi}) are identical. All gauge interactions act
within each of the diagonal blocks corresponding to the irreducible
representations $\rho_1$ and $\rho_2$,
while the off-diagonal block contains the physics that causes mixing
via the mass matrices.

For instance, with this interpretation in mind,
it seems natural to assume the diagonal blocks to be the same for each
generation. This is equivalent to saying that if electroweak interactions 
were switched off quark masses in a given charge sector
would be degenerate. As shown in \cite{CES} and \cite{HAS}, in the case
of two generations, this assumption fixes the Cabibbo angle in terms of
the quark masses. With $m_1,\mu_1$ denoting the masses of the first
generation, say $m_1\equiv m_u$ and $\mu_1\equiv m_d$, and
$m_2,\mu_2$ denoting the masses of the second generation, say
$m_2\equiv m_c$ and $\mu_2\equiv m_s$, one obtains
\bg
\cos\theta =\frac{\sqrt{m_1\mu_1}+\sqrt{m_2\mu_2}}
                 {\sqrt{(m_1+m_2)(\mu_1+\mu_2)}}\, .\label{eq-th}
\eg
In the case of three generations, as worked out in \cite{HAS}, the same
assumption leads to an analytic expression of the CKM matrix (\ref{eq-CKM})
in terms of the quark masses and a few parameters which has a remarkable
similarity to established phenomenological forms \cite{Fri}.\\[12pt]
\newpage\noindent
{\bf 4. Shifting diagonalization to one sector only}

In computing the CKM matrix (\ref{eq-CKM}) one may proceed by independent
diagonalization of the triangular matrices $T^{(u)}$ and $T^{(d)}$ of
eq.~(\ref{eq-mass}), by means of bi-unitary transformations in each
charge sector, cf. eq.~(\ref{eq-biun}). As an alternative to this tedious
calculation \cite{HAS} we now show that diagonalization can be shifted to
one of the charge sectors only, whose effective mass matrix again has
triangular form. This is the content of the following theorem.\\[2pt]
\noindent
{\bf Shift theorem:} {\it Given two nonsingular triangle matrices
relating right-chiral to left-chiral fermion fields,
$T^{(q_1)}$ and $T^{(q_2)}$, in the charge sectors $q_1$ and
$q_2=q_1\pm 1$, respectively, and a bi-unitary transformation which
diagonalizes $T^{(q_1)}$}
\bg
V_L^{(q_1)}T^{(q_1)}V_R^{(q_1)\,\dagger}=\; \stackrel{0}{T}{}^{(q_1)}\, .
\label{eq-TQ1}
\eg
{\it If the same bi-unitary transformation is simultaneously
applied to $T^{(q_2)}$, the charge changing current relating the
left-chiral fermion fields remains
unchanged. Furthermore, by an additional unitary transformation of
the right-chiral fields of charge $q_2$, the transformed matrix can
again be cast into triangular form,}
\bg
V_L^{(q_1)}T^{(q_2)}V_R^{(q_1)\,\dagger}={\cal T}{\cal U}\, .
\label{eq-TT}
\eg
The first part of the statement is fairly obvious: if the same
transformation $V_L^{(q_1)}$ is applied to the left-chiral fields
of both charges, the matrix elements
$\overline{\psi^{(q_1)}}I_\pm\psi^{(q_2)}$ of isospin raising and
lowering operators do not change. The second
part is a consequence of the decomposition theorem. Here, in fact,
the r.h. factor $V_R^{(q_1)\,\dagger}$ is irrelevant because it may
be absorbed in the unitary $\cal U$ acting on the right-chiral fields.
Note that the right factor of the
bi-unitary transformation (\ref{eq-TQ1}) follows from the left factor,
\[
V_R^{(q_1)}=\; \stackrel{0}{T}{}^{(q_1)}
V_L^{(q_1)}\left( T^{(q_1)\, -1}\right)^\dagger\, .
\]

We illustrate the shift theorem for the example of two quark generations
for which the basis (\ref{eq-psi}) reduces to
\[
\Psi =\left(
      u_L^\prime ,d_L^{\,\prime},u_R^\prime,d_R^{\,\prime},
      c_L^\prime ,s_L^\prime ,c_R^\prime ,s_R^\prime
      \right)^T\, .
\]
The mass matrices are
\[
T^{(u)}=\left(\begin{array}{cc}
        \alpha^{(u)}& 0 \\ \kappa^{(u)}& \beta^{(u)}
        \end{array}\right) \, ,\qquad
T^{(d)}=\left(\begin{array}{cc}
        \alpha^{(d)} & 0 \\ \kappa^{(d)} & \beta^{(d)}
        \end{array}\right)\,
\]
where the parameters may be chosen real, without loss of generality.
In this case they are related to the quark masses by
\[
\alpha^{(u)}\beta^{(u)}=m_um_c\, ,\quad 
\alpha^{(u)\, 2}+\beta^{(u)\, 2}+\kappa^{(u)\, 2}=m_u^2+m_c^2
\]
(and analogous relations for the parameters $\alpha^{(d)}$, $\beta^{(d)}$,
$\kappa^{(d)}$, in terms of $m_d$ and $m_s$). Then
\[
V_L^{(u)}=\left(\begin{array}{cc}
          a & -b \\ b & a
          \end{array}\right)\, ,\qquad
V_R^{(u)}=\frac{1}{\alpha^{(u)}}
          \left(\begin{array}{cc}
          m_ua & -m_cb \\ -m_cb & -m_ua
          \end{array}\right)\, ,
\]
where $a$ and $b$ are given by
\[
a=\sqrt{\frac{m_c^2-\alpha^{(u)\, 2}}{m_c^2-m_u^2}}\, ,\qquad b=\sqrt{1-a^2}\, .
\]
(We note in passing that signs were chosen such that
$\stackrel{0}{T}{}^{(u)}=\diag (m_u,-m_c)$.)
In the basis given above the step operators of weak isospin are
represented by
\[
\widehat{I}_\pm =\frac{1}{2}\left( \begin{array}{cc|cc}
       \tau_\pm & 0 & 0 & 0 \\ 0 & 0 & 0 & 0  \\ \hline
       0 & 0 & \tau_\pm & 0 \\ 0 & 0 & 0 & 0
       \end{array}\right)\, ,
\]
where the entries are $2\times 2$ block matrices, $\tau_i$ being Pauli
matrices. The transformation $V_L^{(u)}$, when applied to both charge 
sectors, reads in this basis,
\[
\widehat{V}_L^{(u)}=
       \left( \begin{array}{cc|cc}
       a\id & 0 & -b\id & 0 \\ 0 & \id & 0 & 0  \\ \hline
       b\id & 0 & a\id & 0 \\ 0 & 0 & 0 & \id
       \end{array}\right)\, ,
\]
which obviously commutes with $\widehat{I}_\pm$.

To continue with this example let us introduce the assumption, mentioned
above, of choosing the diagonal blocks to be the same in the two
generations. This implies $\alpha^{(u)}=\beta^{(u)}=\sqrt{m_um_c}$,
$\kappa^{(u)}=m_c-m_u$, $\alpha^{(d)} =\beta^{(d)} =\sqrt{m_dm_s}$, and
$\kappa^{(d)} =m_s-m_d$. The matrix elements ${\cal T}=\{ t_{ik}\}$ are
then found to be
\bg
t_{11}=\frac{1}{\sqrt{m_u+m_c}}\left\{
       \sqrt{m_dm_s}\,\Delta\Sigma +m_d\sqrt{m_um_d}\,\Sigma -
       m_s\sqrt{m_um_s}\,\Delta\right\}^{1/2}\, ,\label{eq-t11}
\eg
\bg
t_{21}=\frac{m_s-m_d}{m_u+m_c}\Delta\Sigma /t_{11}\, ,\quad
t_{22}=-m_dm_s/t_{11}\, .\label{eq-t212}
\eg
In these formulae we have fixed phases such that all entries are real
and that $t_{22}$ is negative. The symbols $\Delta$ and $\Sigma$
stand for
\bg
\Sigma =\sqrt{m_cm_s}+\sqrt{m_um_d}\, ,\qquad
\Delta =\sqrt{m_cm_d}-\sqrt{m_um_s}\, .\label{eq-sigdel}
\eg
It is now straigthforward to determine the single unitary matrix which
diagonalizes the matrix 
$\left( {\cal T}{\cal T}^\dagger\right)$ and to confirm that this is
the Cabibbo matrix with $\theta$ as given by eq.~(\ref{eq-th}). One finds
\[
V^{\mbox{\footnotesize (C)}}=\frac{1}{N}\left(
    \begin{array}{cc}
    \Sigma & -\Delta\\
    \Delta & \Sigma
    \end{array}\right)\, ,
\]
where $N=\sqrt{(m_u + m_c )(m_d + m_s )}$, $\Sigma$ and $\Delta$ being defined
in eq.~(\ref{eq-sigdel}) above. This result is equivalent to the formula
(\ref{eq-th}).

Clearly, the procedure is symmetric in the two charge sectors. The
diagonalization may as well be shifted to the charge sector $(q_1)$.
In the example given this is equivalent to interchanging
\[
m_u \longleftrightarrow m_d\, ,\quad m_c \longleftrightarrow m_s\, .\]
\noindent
{\bf 5. The case of three generations}

Suppose that in the case of three generations we shift diagonalization to
the {\it up\/}- or the {\it down\/}-sector, 
as described in the shift theorem and in
eq.~(\ref{eq-TT}). For the sake of clarity we write the effective mass
matrix ${\cal T}=\{ t_{ik} \}$ in terms of moduli and phases of its entries
as follows
\bg
{\cal T}=\left( \begin{array}{ccc}
          t_{11} & 0 & 0 \\ t_{21} & t_{22} & 0 \\ t_{31} & t_{32} & t_{33}
         \end{array} \right) \equiv
         \left( \begin{array}{ccc}
         \alpha e^{i\varphi_\alpha} & 0 & 0 \\
         \kappa_1 e^{i\varphi_1} & \beta e^{i\varphi_\beta } & 0 \\
         \kappa_3 e^{i\varphi_3} & \kappa_2 e^{i\varphi_2} & \gamma 
e^{i\varphi_\gamma}
         \end{array} \right)\, , \label{eq-Tpar}
\eg
a notation that is consistent with the one employed in the example 
discussed in sec. 4.

For definiteness let us shift the analysis to the {\it down\/}-sector in which
case the matrix $\left( {\cal T}{\cal T}^\dagger\right)$ has eigenvalues 
$\{ m_d^2,m_s^2,m_b^2\}$. From its characteristic polynomial we obtain the 
following equations
\bg
m_d^2m_s^2m_b^2 = \alpha^2\beta^2\gamma^2\, , \label{eq-pol1}
\eg
\begin{eqnarray}
m_d^2m_s^2+m_d^2m_b^2+m_s^2m_b^2 &=& \alpha^2\beta^2+\beta^2\gamma^2+
   \gamma^2\alpha^2+\alpha^2\kappa_2^2+
   \nonumber\\
   & &\beta^2\kappa_3^2+\gamma^2\kappa_1^2+\kappa_1^2\kappa_2^2-
   \nonumber\\
   & &2\beta\kappa_1\kappa_2\kappa_3
       \cos (\varphi_\beta-\varphi_1+\varphi_3-\varphi_2)\, ,
\label{eq-pol2}
\end{eqnarray}
\bg
m_d^2+m_s^2+m_b^2 = \alpha^2+\beta^2+\gamma^2+\kappa_1^2
                   +\kappa_2^2+\kappa_3^2\, .\label{eq-pol3}
\eg
The decomposition theorem (\ref{eq-MTU}) tells us that the matrix $\cal T$,
eq.~(\ref{eq-TT}), is determined up to multiplication by 
$W=\diag (e^{i\omega_1},e^{i\omega_2},e^{i\omega_3})$ from the right, where
the $\omega_i$ are arbitrary. Under this substitution relations
(\ref{eq-pol1}) and (\ref{eq-pol3}) are trivially invariant, relation
(\ref{eq-pol2}) is also invariant because the argument of the cosine is
unchanged. 

Having shifted the diagonalization to the {\it down\/}-sector, the
CKM matrix (\ref{eq-CKM}) is given by the hermitean conjugate of the unitary
matrix that diagonalizes $\left( {\cal T}{\cal T}^\dagger\right)$. Adapting
our earlier results, cf. ref. \cite{HAS}, to the present situation,
we obtain the following analytic
expressions for the CKM matrix in terms of the entries of $\cal T$, i.e.
of the parameters $\alpha ,\ldots ,\varphi_\gamma$ of eq.~(\ref{eq-Tpar}).
\begin{eqnarray}
V^{\mbox{\footnotesize (CKM)}}&\equiv &\left( \begin{array}{ccc}
      Ae^{i\phi_A} & Be^{i\phi_B} & Ce^{i\phi_C}\\
      De^{i\phi_D} & Ee^{i\phi_E} & Fe^{i\phi_F}\\
      Ge^{i\phi_G} & He^{i\phi_H} & Ie^{i\phi_I}
      \end{array}\right) \label{eq-VA} \\ %
      &=&\left( \begin{array}{ccc}
      f(m_d)/N_d & f(m_s)/N_s & f(m_b)/N_b \\
      g(m_d)/N_d & g(m_s)/N_s & g(m_b)/N_b \\
      h(m_d)/N_d & h(m_s)/N_s & h(m_b)/N_b 
      \end{array}\right)\, , \label{eq-V}
\end{eqnarray}
where the functions $f$, $g$, $h$, and the normalization factors are given by
\bg
f(m_i)=\alpha\beta\kappa_1\kappa_2e^{-i(\varphi_1+\varphi_2-
       \varphi_\alpha- \varphi_\beta)}-
       \alpha\kappa_3(\beta^2-m_i^2)e^{-i(\varphi_3-\varphi_\alpha )}\, ,
       \label{eq-f}
\eg
\bg
g(m_i)=m_i^2\kappa_1\kappa_3e^{-i(\varphi_3-\varphi_1)}-
       \beta\kappa_2(\alpha^2-m_i^2)e^{-i(\varphi_2-\varphi_\beta )}\, ,\   
       \label{eq-g}
\eg
\bg
h(m_i)=(\alpha^2-m_i^2)(\beta^2-m_i^2)-\kappa_1^2m_i^2\, ,\label{eq-h}
\eg
with $m_i=m_d$, or $m_s$, or $m_b$,
\bg
N_d=\left\{\left[
    (\alpha^2-m_d^2)(\beta^2-m_d^2)-m_d^2\kappa_1^2\right]
    (m_b^2-m_d^2)(m_s^2-m_d^2)\right\}^{1/2}\, , \label{eq-N}
\eg
and with $N_s$ and $N_b$ obtained from eq.~(\ref{eq-N}) by cyclic 
permutation of $(m_d,m_s,m_b)$. The first eq.~(\ref{eq-VA}) is only meant to
express the matrix elements of 
$V^{\mbox{\footnotesize (CKM)}}$ in terms of their moduli and 
their phases while eq.~(\ref{eq-V}) gives our explicit results in terms
of $\cal T$. So, for instance, reality and sign of eq.~(\ref{eq-h}) implies
$\phi_G=\phi_I=0$, and $\phi_H=\pi$.
Of course, the results fulfill all relations such as $C^2=1-A^2-B^2$ etc.
which follow from unitarity. 

Finally, we recall that one may equally well shift diagonalization
to the {\it up\/}-sector in which case $(m_d,m_s,m_b)$ are replaced by
$(m_u,m_c,m_t)$, while the parameters in eq.~(\ref{eq-Tpar}) take different
values because in the determining equation (\ref{eq-TT}) the charge sectors
are interchanged. In this case the CKM matrix (\ref{eq-CKM}) is given by the
unitary matrix that diagonalizes $\left( {\cal T}{\cal T}^\dagger\right)$
(not its hermitean conjugate).\\[12pt]
{\bf 6. The CP-measure as a function of observables}

As is well known the following nine quantities are rephasing invariants,
i.e. are independent of the specific parametrization of the CKM matrix
one  chooses \cite{DWU}, 
\[
\Delta_{i\alpha}=V^{\mbox{\footnotesize (CKM)}}_{j\beta}
                 V^{\mbox{\footnotesize (CKM)}}_{k\gamma}
                 V^{\mbox{\footnotesize (CKM)}\star}_{j\gamma}
                 V^{\mbox{\footnotesize (CKM)}\star}_{k\beta},\,
   \{ i,j,k\} , \{\alpha ,\beta ,\gamma\}\in \{ 1,2,3\}\mbox{ cyclic}\, .
\]
In particular, unitarity of 
$V^{\mbox{\footnotesize (CKM)}}$ implies that they all have the same 
imaginary part, cf. \cite{ChK}, \cite{Jar}, \cite{Gre},
\bg
{\cal J}:= \mbox{Im }\Delta_{i\alpha}\, , \label{eq-J}
\eg
which is a parametrization independent measure of the amount of CP violation
in the standard model with three generations. We find it useful
to call this quantity
the {\it CP-measure.\/} Our aim in this section is to express the
CP-measure in terms of observables only. For that purpose we start from,
say,
\[
\Delta_{13}=V^{\mbox{\footnotesize (CKM)}}_{21}
            V^{\mbox{\footnotesize (CKM)}}_{32}
            V^{\mbox{\footnotesize (CKM)}\star}_{22}
            V^{\mbox{\footnotesize (CKM)}\star}_{31}=
            -DEGHe^{i(\phi_D-\phi_E)}\, ,
\]
from which
\bg
{\cal J}=\mbox{Im }\Delta_{13}=DEGH\sin (\phi_E -\phi_D)\, .\label{eq-?}
\eg
The calculation of $\sin (\phi_E -\phi_D)$ in terms of the observable
moduli $A,\ldots $ is straightforward but tedious and we refer to the
appendix for an outline of that calculation. The result for the CP-measure
reads\footnote{In principle, the CP-measure is plus or minus the expression
on the r. h. s. of eq.~(\ref{eq-Jar}). The data seems to indicate that 
$\cal J$ is positive, hence our choice of this sign.}
\bg
{\cal J}=\frac{1}{2}\left\{
         4A^2B^2D^2E^2-\left[
         A^2E^2+B^2D^2-(A^2+B^2+D^2+E^2)+1\right]^2
         \right\}^{1/2}\, .                 \label{eq-Jar}
\eg

To the best of our knowledge this formula for the CP-measure is new. It 
expresses the strength of CP violation in terms of moduli of CKM matrix
elements, i.e. in terms of the observable quantities
$A\equiv\vert V_{ud}\vert$, $B\equiv\vert V_{us}\vert$,
$D\equiv\vert V_{cd}\vert$, and $E\equiv\vert V_{cs}\vert$. We have obtained it
in our general framework but it may of course be verified in any
specific parametrization of the CKM-matrix. 

{}From eq.~(\ref{eq-Jar}) the
following symmetries of $\cal J$ are evident:
\begin{itemize}
\item[{}(i)] $\cal J$ is invariant under the exchange 
$B\longleftrightarrow D$.
This property reflects our earlier remark that diagonalization may 
equivalently be shifted to the {\it up\/}-sector in which case the
CKM-matrix equals the diagonalization matrix, not its hermitean conjugate.
\item[(ii)] Simultaneous interchange $A\longleftrightarrow B$ and
$D\longleftrightarrow E$ leaves $\cal J$ invariant.
\end{itemize}
Combining the symmetries (i) and (ii) one sees that the simultaneous
interchange $A\longleftrightarrow D$ and $B\longleftrightarrow E$ is also a
symmetry. Finally, by combining all three of these one shows that
$\cal J$ is also invariant under $A\longleftrightarrow E$.

It easy to verify that $\cal J$ vanishes, as it should,  whenever one of 
the three generations decouples from the other two. For example, if the
first generation decouples, we have $A=1$, hence $B=D=0$ and ${\cal J}=0$.

For given values of the moduli $A$, $B$, and $E$, the CP-measure is
defined only for values of the modulus $D$ in the interval
$(D_1,D_2)$, where
\bg
D_{1,2}=\left\{ ABE\mp\sqrt{(1-A^2-B^2)(1-B^2-E^2)}\right\}/(1-B^2)\, .
\label{eq-D12}
\eg
$\cal J$ vanishes at these boundary points. It assumes its maximal value
at
\bg
D_0= \sqrt{A^2B^2E^2+(1-A^2-B^2)(1-B^2-E^2)}/(1-B^2) \, ,\label{eq-D0}
\eg
at which point the CP-measure takes the value
\bg
{\cal J}(D_0) =\frac{1}{1-B^2}
               \sqrt{A^2B^2E^2(1-A^2-B^2)(1-B^2-E^2)}\, . \label{eq-J0}
\eg

In these formulae the moduli can be expressed in terms of quark masses
and the parameters of $\cal T$, eq.~(\ref{eq-Tpar}), by means of our
formulae (\ref{eq-V} - \ref{eq-N}) above. Alternatively, they may be taken 
from experiment, as in the following example.

According to the minireport in the Review of Particle Properties
\cite{GKR} an overall best fit to the data allows for values of the
magnitude $A$ of the matrix element $V_{ud}$ 
between 0.9745 and 0.9757, i.e. within
an interval of width 0.0012. Similarly, $B$ lies between
0.219 and 0.224, $D$ lies between 0.218 and 0.224, and
$E$ lies between 0.9736 and 0.9750. The
uncertainty of $D$ being the largest was the reason why we solved
our formula (\ref{eq-Jar}) for the CP-measure in terms of that quantity.
Evidently, any other choice is possible. It is amusing to note that
if we take the central values provided by the best fit, 
i.e. $A=0.9751$, $B=0.2215$, $E=0.9743$, we obtain $D_0=0.2213$ for the
point at which $\cal J$ is maximal, cf. eq.~(\ref{eq-D0}), a value that
happens to fall in the center of the allowed interval for D. \\[12pt]
{\bf 7. Reconstruction of the effective mass matrix}

In this section we show how to reconstruct the effective, triangular mass
matrix $\cal T$, eq.~(\ref{eq-Tpar}), from the CKM mixing matrix. This
reconstruction is unique, except for trivial redefinitions of phases
irrelevant for physics, because the triangular form of the effective mass
matrix contains no redundant information. For the sake of definiteness
we again assume that diagonalization is shifted, by the shift 
theorem~(\ref{eq-TT}), to the {\it down\/}-sector, such that the {\it up\/} 
mass sector is already diagonal while the {\it down\/}-sector has the 
effective, triangular form~(\ref{eq-Tpar}). We repeat, however, that the
procedure is completely symmetric in the two charge sectors, and that the
case of a nondiagonal, effective mass matrix in the 
{\it up\/}-sector is obtained from our formulae by simple and obvious
modifications. 

By absorbing redundant phases into the base states , see \cite{HAS}, one
finds that the physically relevant information coded by $\cal T$ is
contained in seven real parameters, viz.
\bg \label{eq-Para}
\alpha\, ,\; \beta\, ,\; \gamma\, ,\;\kappa_1\, ,\;
\kappa_2\, ,\; \kappa_3\, ,\;
\Phi=\varphi_\beta -\varphi_1 +\varphi_3-\varphi_2\, ,
\eg
the first six of which can be chosen positive. Making use of
eqs.~(\ref{eq-pol1} -\ref{eq-pol3}) that follow from the characteristic
polynomial, we are left with four parameters. These will be determined from
the CKM matrix as follows. From eqs.~(\ref{eq-TT}), (\ref{eq-biun}), and 
(\ref{eq-CKM}) we have
\begin{eqnarray} \label{eq-dia}
{\cal T}{\cal T}^\dagger & = & V_L^{(u)}V_L^{(d)\,\dagger}\diag
(m_d^2,m_s^2,m_b^2)V_L^{(d)}V_L^{(u)\,\dagger}\nonumber\\
 & = & V^{\mbox{\footnotesize (CKM)}}\diag (m_d^2,m_s^2,m_b^2)
       V^{\mbox{\footnotesize (CKM)}\dagger}\, .
\end{eqnarray}
Denoting the moduli of CKM matrix elements as in eq.~(\ref{eq-VA}) one
derives the following expressions from eq.~(\ref{eq-dia}):
\bg \label{eq-P1}
\alpha^2\;\;\;\;\;\, =\;\; m_d^2A^2+m_s^2B^2+m_b^2(1-A^2-B^2)\, ,
\eg
\begin{eqnarray} \label{eq-P2}
\alpha^2\kappa_1^2 & = & 
    m_d^4A^2D^2+m_s^4B^2E^2+m_b^4(1-A^2-B^2)(1-D^2-E^2)+\nonumber\\
& & (m_d^2m_s^2-m_d^2m_b^2-m_s^2m_b^2)
    ( 1-A^2-B^2-D^2-E^2+A^2E^2+B^2D^2) -\nonumber\\
& & 2m_b^2( m_d^2A^2D^2+m_s^2B^2E^2) \, ,
\end{eqnarray}
\begin{eqnarray} \label{eq-P3}
\alpha^2\kappa_3^2 & = & m_d^4A^2(1-A^2-D^2)+m_s^4B^2(1-B^2-E^2)+\nonumber\\
& & m_b^4(1-A^2-B^2)(A^2+B^2+D^2+E^2-1)-\nonumber\\
& & (m_d^2m_s^2-m_d^2m_b^2-m_s^2m_b^2)
    (1-A^2-B^2-D^2-E^2+A^2E^2+B^2D^2+2A^2B^2)\nonumber\\
& & +2m_b^2\left[ m_d^2A^2(A^2+D^2-1)+m_s^2B^2(B^2+E^2-1)\right]\, ,
\end{eqnarray}
\bg \label{eq-P4}
\beta^2+\kappa_1^2 =\;\; m_d^2D^2+m_s^2E^2+m_b^2(1-D^2-E^2)\, .
\eg 
These equations, together with eqs.~(\ref{eq-pol1} - \ref{eq-pol3}), are
sufficient to calculate the set (\ref{eq-Para}), once the moduli of the
CKM mixing matrix and the quark masses are given. Thus, we obtain explicit
and unambiguous expressions for the parameters (\ref{eq-Para}) which
determine the effective mass matrix $\cal T$, in terms of observables only.

If one of the charge sectors, say
the {\it up\/}-sector, is diagonal from the start, the problem of 
reconstructing the mass matrix from the data is completely solved.
If the two sectors are treated more symmetrically and if
the mass matrices are nondiagonal in either charge sector,
one might wish to go one step further by
trying to reconstruct the original nondiagonal, triangular mass matrices
$T^{(u)}$ and $T^{(d)}$ from the effective matrix $\cal T$. A promising
example would be the physically interesting case mentioned above, 
where these matrices have equal entries in the main diagonal. Although 
we have the necessary analytic formulae at our disposal, 
cf. eqs.~(\ref{eq-t1} - \ref{eq-t3}), this reconstruction 
is rather lengthy and tedious, and we leave it to later investigation.

The CKM observables and the quark masses have appreciable experimental 
uncertainties which will 
determine present error bars of the parameters (\ref{eq-Para}). In a future
publication we intend to perform a detailed numerical analysis including
an estimate of errors. \\[12pt]
{\bf 8. Summary and conclusions}

In the minimal standard model right-chiral fields do not participate in the
charged current weak interaction and, as far as the interactions with 
vector bosons is concerned, the model is immune against base transformations 
of right-chiral fields. Making use of this freedom in the choice of bases 
for right-chiral quark fields we showed that every nonsingular mass matrix 
is equivalent to a triangular matrix whose entries are calculated in 
eqs.~(\ref{eq-t1} - \ref{eq-t3}). In contrast to a more general form of
the mass matrix (in a given charge sector), the equivalent triangular form
is optimized in the sense that it eliminates all redundant parameters and
exhibits in a simple and transparent manner the remaining freedom in
choosing unobservable phases.
In fact, the triangular form is suggestive and natural if the quark
generations fall into representations of ``semi-sum'' type, i.e. which
are reducible but indecomposable, cf. eq.~(\ref{eq-semi}). Such
representations are typical for super Lie algebras and have been discussed
in the context of electroweak interactions and non-commutative
geometry \cite{CES,HAS,HPS}. 

We then showed that even if both charge sectors, {\it up\/} and
{\it down\/}, initially have nondiagonal, nonsingular mass sectors, 
diagonalization can be shifted to one charge sector only, the resulting
effective mass matrix having again triangular form. Once the latter is
known, the elements of the CKM matrix can be calculated analytically,
cf. eqs.~(\ref{eq-VA} - \ref{eq-N}). In turn, the effective, triangular
mass matrix is reconstructed analytically from the moduli of the CKM
matrix elements, cf. eqs.~(\ref{eq-dia} - \ref{eq-P4}). This
reconstruction is unique up to trivial phase redefinitions. Because the
procedure is independent of any specific parametrization of the CKM matrix,
but mainly because it is economic, concise and transparent, we strongly
advocate the use of triangular matrices in describing quark and lepton
mass sectors.

Finally, in studying an invariant measure for CP violation, defined in
eq.~(\ref{eq-J}), we derived a formula for this CP-measure in terms of
moduli of the CKM matrix elements, i.e. in terms of observables. To the
best of our knowledge this formula, eq.~(\ref{eq-Jar}), is new.

We illustrated our procedure by simple examples for two and three
generations. A more detailed numerical analysis including the available
experimental data with its error bars is left for a future investigation.
An application to charged lepton and neutrino masses, and the consequences
for neutrino oscillations, is in preparation \cite{HPS}.

%%%%%%%%%%%%%%%%%%%%%%%%%%%%%%%%%%%%%%%%%%%%%%%%%%%%%%%%%%%%%%%%%%%%%%%%%
\newpage\noindent
{\bf Appendix}\\[0.3cm]
In this appendix we present an outline of the calculation of the
CP-measure ${\cal J}$ in terms of observable quantities only. According to
eq.~(\ref{eq-?}) we have to determine sin$(\phi_E - \phi_D )$ as a function of
the moduli appearing in $V^{\mbox{\footnotesize (CKM)}}$. 
To this end we first calculate,
using eqs.~(\ref{eq-VA}) and (\ref{eq-V}),
\begin{eqnarray}
\label{gla.1}
  \frac{E}{D} e^{i(\phi_E - \phi_D )} & = &
  \frac{E e^{i\phi_E}}{D e^{i\phi_D}} \; \; = \; \;
  \frac{g(m_s)}{N_s} \frac{N_d}{g(m_d)} \nonumber \\
  & = & \frac{N_d^2}{|g(m_d)|^2} \frac{1}{N_d N_s} g(m_s) g(m_d)^{\star}
  \; \; ,
\end{eqnarray}
from which we have
\begin{equation}
\label{gla.2}
  \mbox{sin} (\phi_E - \phi_D ) =
  \frac{1}{DE} \frac{1}{N_d N_s} \mbox{Im} [g(m_s) g(m_d)^{\star} ] \; \; .
\end{equation}
The determination of $N_d, N_s$ in terms of observables almost immediately
follows from eqs.~(\ref{eq-h}), (\ref{eq-N}):
\begin{eqnarray}
\label{gla.3}
  N_d & = & G (m_b^2 - m_d^2)(m_s^2 - m_d^2) \\
\label{gla.4}
  N_s & = & H (m_b^2 - m_s^2)(m_s^2 - m_d^2)
\end{eqnarray}
Making use of the explicit expression for the function $g$, eq.~(\ref{eq-g}),
we next derive
\bg
\label{gla.5}
  \mbox{Im} [g(m_s) g(m_d)^{\star} ] =
  (m_s^2 - m_d^2) \alpha^2 \beta \kappa_1 \kappa_2 \kappa_3
  \mbox{sin}(\varphi_{\beta} - \varphi_1 + \varphi_3 - \varphi_2 ) \; \; .
\eg
Collecting our results (\ref{gla.2}) - (\ref{gla.5}) the CP-measure
${\cal J}$ is preliminarily given by
\bg
\label{gla.6}
  {\cal J} = \frac{\alpha^2 \beta \kappa_1 \kappa_2 \kappa_3
                   \mbox{sin}\Phi }{(m_b^2 - m_d^2)
                   (m_b^2 - m_s^2)(m_s^2 - m_d^2)}
\eg
and, therefore, is directly proportional to  
$\Phi \equiv \varphi_{\beta} - \varphi_1 + \varphi_3 - \varphi_2$, 
the only physically relevant phase factor in ${\cal T}$.\\
Finally, we remain with the calculation of $\alpha^2 \beta \kappa_1
\kappa_2 \kappa_3$sin$\Phi$. This calculation is quite lengthy but
nevertheless straightforward and was partially performed with the help of
MATHEMATICA. The best strategy is to start from the squared expression
\begin{displaymath}
  (\alpha^2 \beta \kappa_1 \kappa_2 \kappa_3 \mbox{sin} \Phi )^2 =
  \alpha^4 \beta^2 \kappa_1^2 \kappa_2^2 \kappa_3^2 -
  \alpha^4 \beta^2 \kappa_1^2 \kappa_2^2 \kappa_3^2 \mbox{cos}^2 \Phi
  \; \; ,
\end{displaymath}
to exploit eqs.~(\ref{eq-pol1}) - (\ref{eq-pol3}) in order to eliminate
$\kappa_3^2, \beta \kappa_1 \kappa_2 \kappa_3$cos$\Phi$ and
$\gamma^2$ and then to make use of the formulae
\begin{eqnarray}
\label{gla.7}
  \alpha^2 \kappa_2^2 & = &
   \frac{1}{\beta^2}(m_s^2m_b^2-\alpha^2\beta^2)(\beta^2-m_d^2)
    - A^2 (m_b^2 - m_d^2)(m_s^2 - m_d^2) \nonumber \\
  m_d^2 \kappa_1^2 & = &
    (\alpha^2 - m_d^2)(\beta^2 - m_d^2) -
    (1- A^2 - D^2)(m_b^2 - m_d^2)(m_s^2 - m_d^2) \nonumber \\
  \alpha^2 & = & m_b^2 - A^2 (m_b^2 - m_d^2) -
    B^2 (m_b^2 - m_s^2) \\
  \alpha^2 \beta^2 & = &
    m_s^2 m_b^2 (1 - A^2 - D^2) +
    m_d^2 m_b^2 (1 - B^2 - E^2) +
    m_d^2 m_s^2 (A^2 + B^2 + D^2 + E^2 - 1) \nonumber 
\end{eqnarray}
which follow from eqs.~(\ref{eq-VA}), (\ref{eq-V}). The final answer proving
eq.~(\ref{eq-Jar}) is
\begin{eqnarray}
\label{gla.8}
  \alpha^2 \beta \kappa_1 \kappa_2 \kappa_3 \mbox{sin} \Phi & = &
  \frac{1}{2} (m_b^2 - m_d^2)(m_b^2 - m_s^2)(m_s^2 - m_d^2) \times \\
  & & \! \! \! \! \! \!  \left\{ 4 A^2 B^2 D^2 E^2 -
  [A^2 E^2 + B^2 D^2 - (A^2 + B^2 + D^2 + E^2) + 1 ]^2 \right\}^{1/2} 
  . \nonumber
\end{eqnarray}

%mittelmarke
% ganz unten
%%%%%%%%%%%%%%%%%%%%%%%%%%%%%%%%%%%
%%%%%%%%%%%%%%%%%%%%%%%%%%%%%%%%%%%%%%%%%%%%%%%%%%

\end{document}